\documentclass{aastex63}

\usepackage{multirow}
\usepackage{amsmath}

\begin{document}

\title{GW190521 and the GWTC-1 Events: Implication on the Black Hole Mass Function of Coalescing Binary Black Hole Systems}
\author{Yuan-Zhu Wang}
\affil{Key Laboratory of dark Matter and Space Astronomy, Purple Mountain Observatory, Chinese Academy of Sciences, Nanjing, 210033, People's Republic of China.}
\author{Shao-Peng Tang}
\affil{Key Laboratory of dark Matter and Space Astronomy, Purple Mountain Observatory, Chinese Academy of Sciences, Nanjing, 210033, People's Republic of China.}
\affil{School of Astronomy and Space Science, University of Science and Technology of China, Hefei, Anhui 230026, People's Republic of China.}
\author{Yun-Feng Liang}
\affil{Guangxi Key Laboratory for the Relativistic Astrophysics, Department of Physics, Guangxi University, Nanning 530004, People's Republic of China.}
\author{Ming-Zhe Han}
\affil{Key Laboratory of dark Matter and Space Astronomy, Purple Mountain Observatory, Chinese Academy of Sciences, Nanjing, 210033, People's Republic of China.}
\affil{School of Astronomy and Space Science, University of Science and Technology of China, Hefei, Anhui 230026, People's Republic of China.}
\author{Xiang Li}
\author{Zhi-Ping Jin}
\author{Yi-Zhong Fan}
\author{Da-Ming Wei}
\email{Corresponding authors:~yzfan@pmo.ac.cn (YZF) and dmwei@pmo.ac.cn (DMW)}
\affil{Key Laboratory of dark Matter and Space Astronomy, Purple Mountain Observatory, Chinese Academy of Sciences, Nanjing, 210033, People's Republic of China.}
\affil{School of Astronomy and Space Science, University of Science and Technology of China, Hefei, Anhui 230026, People's Republic of China.}

\begin{abstract}
With the black hole mass function (BHMF; assuming an exponential cutoff at a mass of $\sim 40\,M_\odot$) of coalescing binary black hole systems constructed with the events detected in the O1 run of the advanced LIGO/Virgo network, \citet{2017arXiv170501881L} predicted that the birth of the lightest intermediate mass black holes (LIMBHs; with a final mass of $\gtrsim 100\,M_\odot$) is very likely to be caught by the advanced LIGO/Virgo detectors in their O3 run. The O1 and O2 observation run data, however, strongly favor a cutoff of the BHMF much sharper than the exponential one. In this work we show that a power-law function followed by a sudden drop at $\sim 40\,M_\odot$ by a factor of $\sim $a few tens and then a new power-law component extending to $\geq 100M_\odot$ are consistent with the O1 and O2 observation run data. With this new BHMF, quite a few LIMBH events can be detected in the O3 observation run of advanced LIGO/Virgo. The first LIMBH born in GW190521, an event detected in the early stage of the O3 run of advanced LIGO/Virgo network, provides additional motivation for our hypothesis.
\end{abstract}

\section{Introduction}
The stellar-mass (i.e., $\leq 100\,M_\odot$) black holes (BHs) are expected to form when very massive stars collapse at the end of their life cycle \citep{2002RvMP...74.1015W} or merger/accretion from lighter compact objects. Traditionally, the masses of the BHs can be measured if they are within the binary systems. Indeed, a few dozens of BHs of stellar mass have already been detected in X-ray binaries within the Milky Way and some nearby galaxies \citep{2011ApJ...741..103F}. On 2015 September 14, the successful detection of a gravitational wave (GW) signal from the merger of a binary black hole (BBH) by Advanced LIGO \citep[aLIGO;][]{2016PhRvL.116f1102A} opens a brand-new window into observing the universe. So far, about two dozens of BBH GW events have been reported and even more events are expected to be released soon.

In the population studies, the black hole mass functions (BHMFs) in different binary systems are one of the key subjects since such information can help us reveal the stellar evolution physics and/or the origin of these systems. For instance, there could be the suppression on the number of the low mass BHs and the absence of very massive stellar-mass BHs with masses above $\sim 52-65M_\odot$, as expected in the modern supernova explosion models \citep[e.g.][]{Belczynski2016,2020ApJ...896...56W}. In principle, the BHMFs for different binary systems (for instance, the X-ray binaries, the BBH merger events and the neutron star-BH merger events) may show different characters \citep{2020ApJ...892...56T}. It is therefore essential to construct them with the observation data and some interesting features may be revealed. Before the recent report of a BH candidate with a mass of $3.3^{+2.8}_{-0.7}M_\odot$ identified in the binary system 2MASS J05215658+4359220 \citep{2019Sci...366..637T}, the BHs identified in the X-ray binaries have the lightest mass of $\sim 5M_\odot$, which is much heavier than the upper limit of neutron stars \citep{2010ApJ...725.1918O} and suggests the presence of a mass gap between the neutron stars and the BHs. Since the discovery of the BBH merger event, the GW data have been extensively adopted to reconstruct the BHMF in such a specific group of objects \citep[e.g.,][]{2016PhRvX...6d1015A,2017ApJ...851L..25F, 2017PhRvD..95j3010K,2018arXiv180204909B, 2019ApJ...882L..24A}. While for the O1 events, the BHMF was only loosely constrained due to the rather small sample. Interestingly, assuming an exponential cutoff of the mass function at $\sim 40M_\odot$, the predicted birth rate of the lightest intermediate mass BH (i.e., $M_{\rm f}\geq 100M_\odot$, where $M_{\rm f}$ is the final mass of the new BH formed in the merger) is high and the detection prospect is found to be quite promising in the O3 run of the advanced LIGO/Virgo \citep{2017arXiv170501881L}. Such a result is indeed encouraging. However, the analyses of both O1 and O2 BBH events find a ``termination" of the BHMF at $\sim 40\,M_\odot$ \citep{2017ApJ...851L..25F, 2018arXiv180204909B, 2019MNRAS.484.4216R, 2018ApJ...856..173T, 2019ApJ...882L..24A, 2019PhRvD.100d3012W}, which has been taken as the evidence for the (pulsational) pair instability processes of the supernova explosions. Consequently, the detection prospect of IMBH is unpromising. However, heavier stellar-mass BHs certainly exist in the Universe (for instance in GWTC-1 there were BHs formed with $M_{\rm f}\sim 60-80M_\odot$ \citep{2019PhRvX...9c1040A}; see Sec.\ref{sec:models} for extended discussion of this issue) and they may also be involved in the BBH mergers. We hence speculate the presence of a new component in the ``high" mass range but there is a sudden drop in the BHMF at the mass of $m_{\rm max}$ due to the supernova explosion mechanism limit. Note that {\it in the final stage of preparing for this work}, the LIGO/Virgo collaboration reported the first IMBH formed in GW190521, a merger of two BHs with the masses of $\sim 60-80\,M_\odot$ \citep{2020arXiv200901075T}. These authors also pointed out that the detection of an IMBH in the early stage of O3 run of advanced LIGO/Virgo is in tension with the O1-O2 data \citep{2020arXiv200901190T}. As we will demonstrate in this work, such a tension can be solved in our new BHMF model.

This work is organized as the follows: in Sec.\ref{sec:methods}, we introduce the models for parameterizing the mass distributions, the likelihood of hierarchical inference, and the selection effects. We report our results in Sec.\ref{sec:res} and discuss the implication of predicting the IMBH in O3 run in Sec.\ref{sec:imp}. And Sec.\ref{sec:dis} is our Conclusion and Discussion.

\section{Data, Models and Selection Effects}\label{sec:methods}
\subsection{Parameterized Mass Spectrums}\label{sec:models}
\begin{figure}[]
	\figurenum{1}\label{fig:the2models}
	\centering
	\includegraphics[angle=0,scale=0.5]{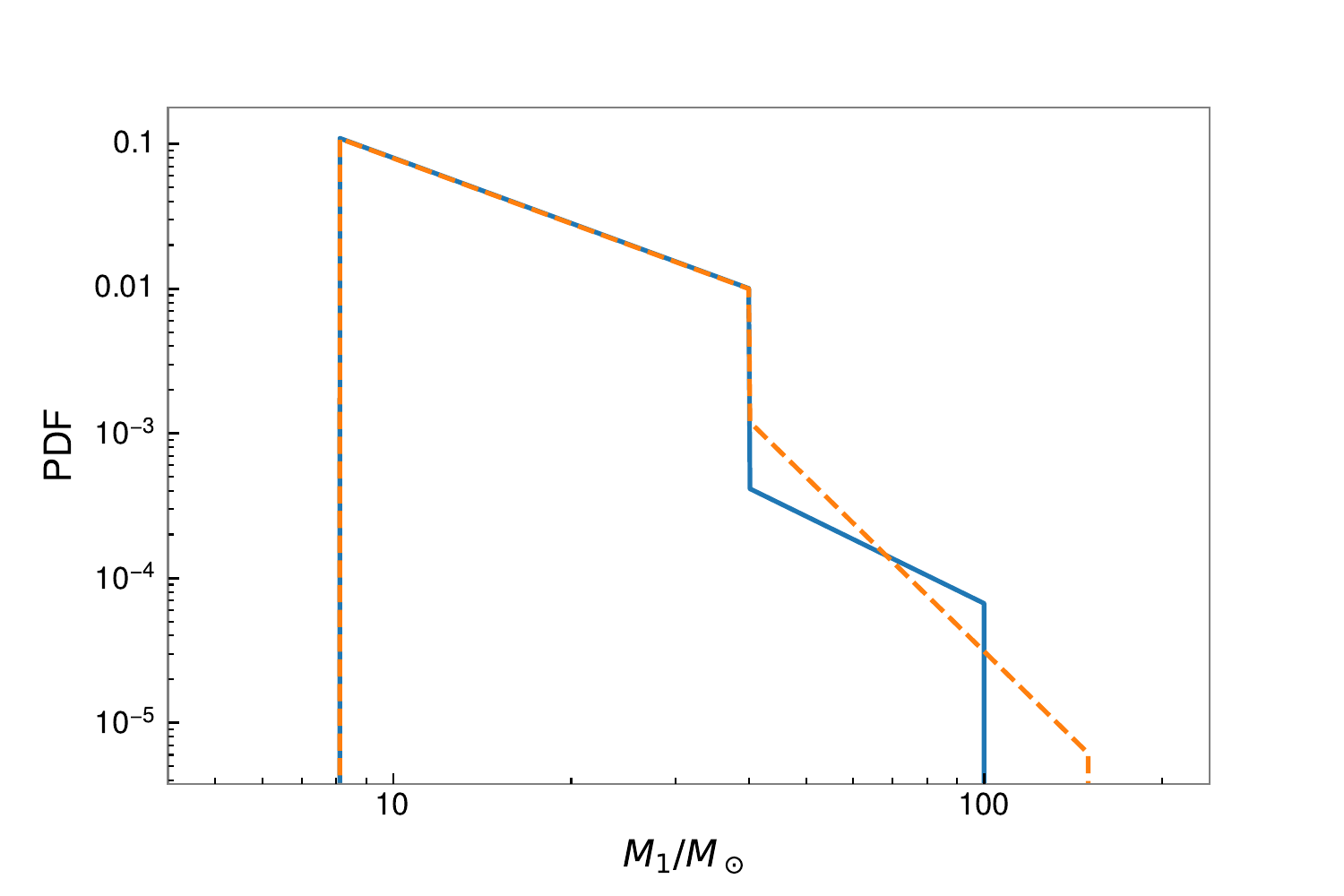}
	\includegraphics[angle=0,scale=0.5]{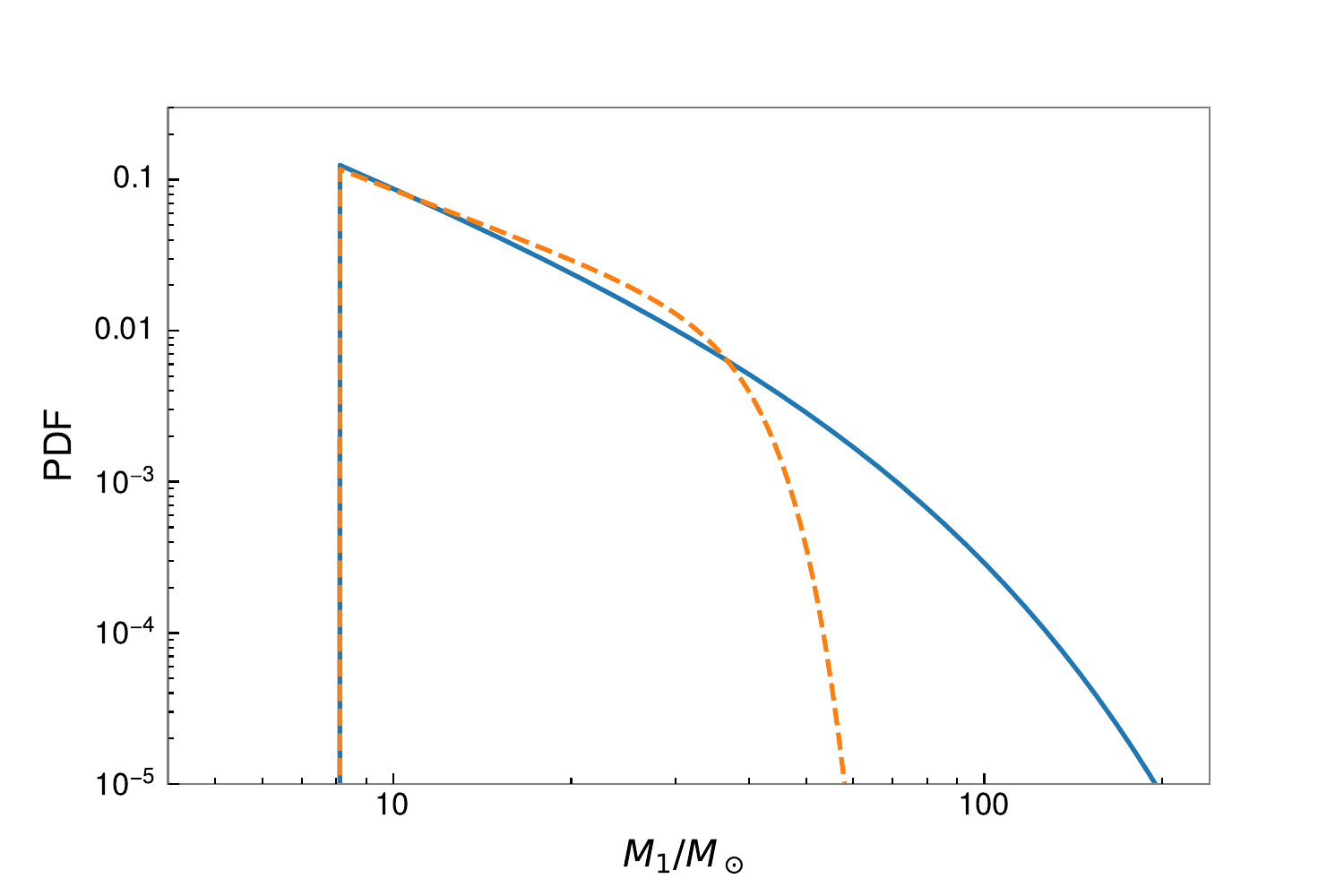}
	\caption{Representative mass distributions of the models. Left panel: the Bcut model, the blue solid line has parameters of $(m_{\rm min}, \alpha, m_{\rm cut}, k)=(8, 1.5, 40, 1)$, and the orange dashed line has the same parameters except $k=5$; right panel: the 2PL model, the blue solid line has parameters of $(m_{\rm min}, m_{\rm max}, m_{\rm edge}, \alpha_1, \alpha_2, {\rm lg}F)=(8, 40, 100, 1.5, 2, -2)$, and the orange dashed line has $(8, 40, 150, 1.5, 4, -1.8)$.}
	\hfill
\end{figure}

We first introduce the BHMF models adopted in \citet[][hereafter LVC19]{2019ApJ...882L..24A}, where three models (namely Model A, B, and C) are discussed. Model A and B share the same formula that are constructed by a power-law (PL) distribution with a hard cutoff at the high mass end,
\begin{equation}
p(m_1,m_2 \mid \Lambda) \propto \begin{cases} N(m_1)\,m_1^{-\alpha}\,q^{\beta_{\rm q}} & \mbox{if }~m_{\rm min} \leq m_2 \leq m_1 \leq m_{\rm max}, \\ 0 &\mbox{otherwise,} \end{cases}
\end{equation}
where $N(m_1)$ is chosen so that the marginal distribution is a power-law in $m_1$. In Model A, $m_{\min}$ and $\beta_{\rm q}$ are fixed to $5$ and $0$ respectively, while in Model B all of the parameters $\Lambda=(\alpha, m_{\rm max}, m_{\rm min}, \beta_{\rm q}$) are allowed to vary. Since Model A of LVC19 is a simplified version of Model B, and there is only mild evidence that the data is better fitted by the complex Model C \citep{2019ApJ...882L..24A}, we only take their Model B as the fiducial model and its name is unchanged to keep the ``consistency" among the literature. To further examine the properties of the mass function around the possible high mass gap, we propose two new BHMFs, i.e., Model Bcut and Model 2PL. Model Bcut is analogue to Model B, but ends smoothly, which is described by
\begin{equation}\label{eq:Mbcut}
p(m_1,m_2 \mid \Lambda) \propto \begin{cases} m_1^{-\alpha}\,{\rm exp}[-(m_1/m_{\rm cut})^{k}]\,q^{\beta_{\rm q}} & \mbox{if }~m_{\rm min} \leq m_2 \leq m_1 \leq m_{\rm max}, \\ 0 &\mbox{otherwise,} \end{cases}
\end{equation}
where the steepness of decline after the cut-off mass $m_{\rm cut}$ is governed by the ``cut-off index" $k$. For $k\leq 1$, there is a good fraction of BHs with masses above $m_{\rm cut}$. While for $k\gg 1$, our case reduces to Model B of LVC19. And the Model 2PL consists of two power-law segments differing greatly in their magnitudes, which reads
\begin{equation}\label{eq:M2pl}
p(m_1,m_2 \mid \Lambda) = \begin{cases} (1-10^{{\rm lg}F})\,A_1\,m_1^{-\alpha_1}\,q^{\beta_{\rm q,1}} & \mbox{if }~m_{\rm min} \leq m_2 \leq m_1 \leq m_{\rm max},
\\ 10^{{\rm lg}F}\,A_2\,m_1^{-\alpha_2}\,q^{\beta_{\rm q,2}} & \mbox{if }~m_{\rm max} \leq m_1 \leq m_{\rm edge} ~ {\rm and } ~m_{\rm min} \leq m_2 \leq m_1, \end{cases}
\end{equation}
where $A_1$ and $A_2$ represent the normalization factor of the first and second segment respectively, and $F$ is the fraction of BHs between the second segment and the whole population. This model is motivated by some astrophysical theories in which the merging black holes can have different origins. For instance, some BBH systems, in particular those reside within the accretion disks of the Active Galactic Nuclei (AGN) \citep{2017ApJ...835..165B, 2017MNRAS.464..946S, 2018ApJ...866...66M} may be able to accrete material from the surrounding and have higher masses. It has also been proposed that heavy black holes can be dynamically formed by lighter objects through hierarchical merger or through runaway collisions \citep{Rodriguez2015,2016ApJ...831..187A, 2016MNRAS.459.3432M, 2017ApJ...840L..24F}. For example, the BBH system of GW170729 may be formed through hierarchical mergers in the migration traps that developed in the accretion disks of AGN \citep{2019PhRvL.123r1101Y}. So it is reasonable to expect an extended tail of the mass distribution of the BHs or a population of ``high" mass objects following a sudden drop of the BHMF at $m_{\rm max}$. Since only GW170729 has a primary mass with inferred median value exceeding $40\,M_\odot$ during the O1 and O2 runs, we expect that the second segment cannot be well constrained from current data. Thus we only take the fixed values of $\beta_2=0$, $\alpha_2=1~({\rm or}~2)$, $m_{\rm edge}=100~({\rm or}~150)$, and a series of ${\rm lg}F$ ranging from $-3$ to $-1$ into consideration in our analysis. To illustrate Model Bcut and Model 2PL more clearly, we present the representative distributions of the primary mass with arbitrary choice of parameter values in Fig.\ref{fig:the2models}. All of our Models and the corresponding priors are summarized in Tab.\ref{tb:priors}.

\begin{table*}[]
	\centering
	\caption{Priors of the Parameters for Different Models}
	\begin{ruledtabular}
		\label{tb:priors}
		\begin{tabular}{c|ccc}
			Parameters/Models                                 & Model B    & Model Bcut   & Model 2PL       \\ \hline
			$\alpha$ or $\alpha_1$                            & [-4, 12]   & [-4, 12]     & [-4, 12]        \\
			$m_{\rm max}$ [$M_\odot$]                                    & [20, 100]  & N/A          & [20, 100]       \\
			$m_{\rm min}$ [$M_\odot$]                                    & [5, 10]    & [5, 10]      & [5, 10]         \\
			$\beta_{\rm q}$ or $\beta_{\rm q,1}$              & [-4, 12]   & [-4, 12]     & [-4, 12]        \\
			$m_{\rm cut}$ [$M_\odot$]                                    & N/A        & [20, 100]    & N/A             \\
			$k$                                               & N/A        & [0, 12]      & N/A             \\
			$\alpha_2$                                        & N/A        & N/A          & 1 and 2         \\
			$\beta_{\rm q,2}$                                 & N/A        & N/A          & 0               \\
			$m_{\rm edge}$ [$M_\odot$]                                   & N/A        & N/A          & 100 and 150     \\
			${\rm lg}F$                                       & N/A        & N/A          & fixed values    \\
		\end{tabular}
		\tablenotetext{}{1. All of the parameters are uniformly distributed in their domains. We have also used a narrower prior volume to calculate the Bayes factors discussed in Sec.\ref{sec:res}, which is $m_{\rm max} \in {\rm U}(30,50)$, $m_{\rm min} \in {\rm U}(6,10)$,$\alpha \in {\rm U}(0,12)$ and $\beta \in {\rm U}(0,12)$. \\ 2. The ${\rm lg}F$ in Model 2PL are fixed to $-3,-2.5,-2,-1.8,-1.5,-1.2$ and $-1$.}
	\end{ruledtabular}
\end{table*}

\subsection{The Likelihood and Selection Effects}
The likelihood for hierarchical inference is constructed based on \citet{2019PASA...36...10T}. For a series of measurements of N events $\vec{d}$, the likelihood for the hyperparameter $\Lambda$ can be inferred via
\begin{equation}\label{eq:llh}
\mathcal{L}(\vec{d}\mid \Lambda) = \prod_{i}^{N}\frac{\mathcal{Z}_{\varnothing}(d_i)}{n_i}\sum_{k}^{n_i}\frac{\pi(\theta_{i}^k\mid \Lambda)}{\pi(\theta_{i}^k\mid \varnothing)}.
\end{equation}
In Eq.(\ref{eq:llh}), the $n_i$ posterior samples for the $i$-th event, the evidence $\mathcal{Z}_{\varnothing}(d_i)$ as well as the default prior $\pi(\theta_k \mid \varnothing)$ are available for the released O1 and O2 events\footnote{\url{https://dcc.ligo.org/LIGO-P2000193/public}}. Taking the selection effect into account \citep{2019PASA...36...10T}, the likelihood can be revised to
\begin{equation}\label{eq:select_llh}
\mathcal{L}(d, N\mid \Lambda,det) = \begin{cases} \frac{1}{p_{\rm det}(\Lambda \mid N)}\mathcal{L}(d, N\mid \Lambda,R) & \mbox{if }~\rho \geq \rho_{\rm th}, \\ 0 & \mbox{otherwise,} \end{cases}
\end{equation}
where $p_{\rm det}(\Lambda \mid N)$ is the probability of detecting $N$ events, which is dependent of population hyperparameter $\Lambda$. If we assume that the Poisson-distributed rate $R$ has a uniform-in-log prior, the marginalized $p_{\rm det}(\Lambda \mid N)$ \citep{2019PASA...36...10T} can be expressed as
\begin{equation}\label{eq:pdet}
p_{\rm det}(\Lambda \mid N) \propto \left( \frac{\mathcal{V}(\Lambda)}{\mathcal{V}_{\rm tot}} \right)^N = f(\Lambda)^N,
\end{equation}
where $\mathcal{V}_{\rm tot}$ is the total spacetime volume and $\mathcal{V}(\Lambda)$ is the visible volume of the population described by $\Lambda$, and $f(\Lambda)$ is the fraction of detectable sources.

The $f(\Lambda)$ for specific hyperparameters $\Lambda$ can be obtained by carrying out injection campaign into the searching pipeline, however it is very computational expensive especially for models with high dimensional parameter space. \citet{2016PhRvX...6d1015A} points out that the searching process can be approximated by a semi-analytic method, which assumes that a source is detectable when the signal-to-noise ratio (SNR) produced in a single detector $\geq 8$ \citep{2019ApJ...882L..24A}. The SNR of an event can be calculated \citep{2020arXiv200705579V} by
\begin{equation}\label{eq:snr}
\rho(\theta)=w({\rm R.A.},{\rm dec.},\iota,\psi)\,\rho_{\rm oo}(m_1,m_2,z),
\end{equation}
where $\rho_{\rm oo}$ is the SNR of an optimally oriented (i.e., faced-on/directly overhead) source with component masses $m_{1, 2}$ at redshift $z$, and $w$ is the angular factor \citep{1993PhRvD..47.2198F} accounting for different sky position and orientation. The distribution of $w$ is available in \citep{1993PhRvD..47.2198F}, thus $f(\Lambda)$ can be evaluated \citep{2020arXiv200705579V} by
\begin{equation}\label{eq:frac}
f(\Lambda) = \int \,{\rm d}m_1\,{\rm d}m_2\,{\rm d}z\,{\rm CCDF}_w\!\left(\frac{\rho_{\rm thr}}{\rho_{\rm oo}}\right)\,\pi(m_1,m_2,z \mid \Lambda),
\end{equation}
where ${\rm CCDF}_w$ denotes the complementary cumulative distribution function of $w$, and $\rho_{\rm thr}$ is chosen to be $8$. To implement fast generating of $\rho_{\rm oo}(m_1,m_2,z)$, we first calculate the optimal SNRs for a series of sources with detector frame masses $(m_1^{\rm det},m_2^{\rm det})$ (yielded from a regular grid) and a reference luminosity distance $d_{\rm L}^{\rm ref}$. The {\sc IMRPhenomPv2} waveform model \citep{2014PhRvL.113o1101H, 2019PhRvD.100b4059K} and the {\it aLIGOEarlyHighSensitivity} PSD taken from {\sc PyCBC} (a good approximation to the PSD during the O1 and O2 runs \citep{2017ApJ...851L..25F}) are used in the calculation. Then $\rho_{\rm oo}(m_1,m_2,z)$ can be derived by interpolating the results of the above calculation and rescaling to a specific distance,
\begin{equation}
\rho_{\rm oo}(m_1,m_2,z) = \rho_{\rm ref}(m_1^{\rm det}, m_2^{\rm det}, d_{\rm L}^{\rm ref})\times \frac{d_{\rm L}^{\rm ref}}{d_{\rm L}(z)}.
\end{equation}
Therefore, the $f(\Lambda)$ in Eq.(\ref{eq:frac}) can be obtained by using Monte-Carlo integration. Finally, we use the python package {\sc Bilby} and {\sc PyMultinest} sampler to obtain the Bayesian evidence and posteriors of the parameters for each models.

\section{Results}\label{sec:res}
\begin{figure}[]
	\figurenum{2}\label{fig:depend}
	\centering
	\includegraphics[angle=0,scale=0.5]{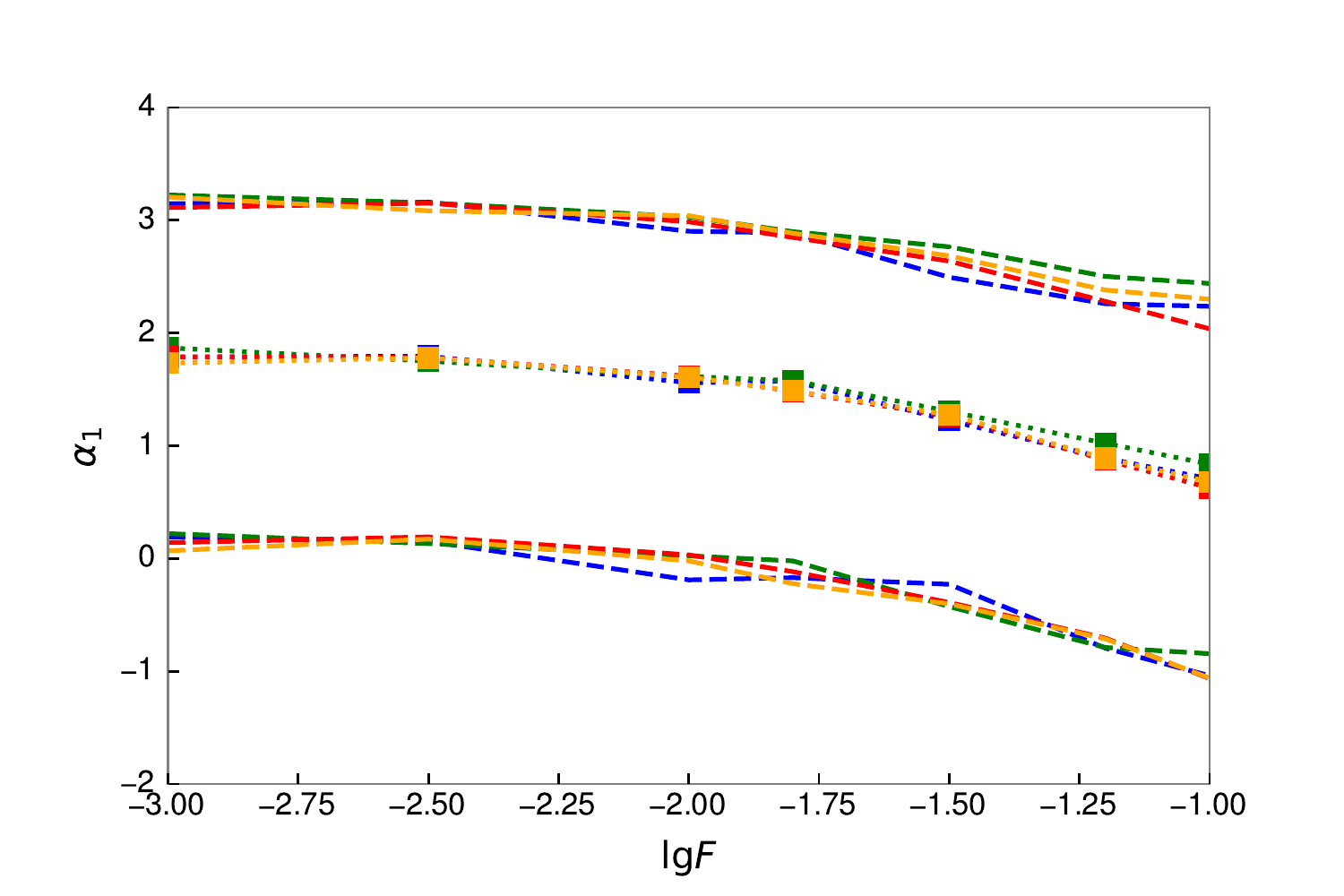}
	\includegraphics[angle=0,scale=0.5]{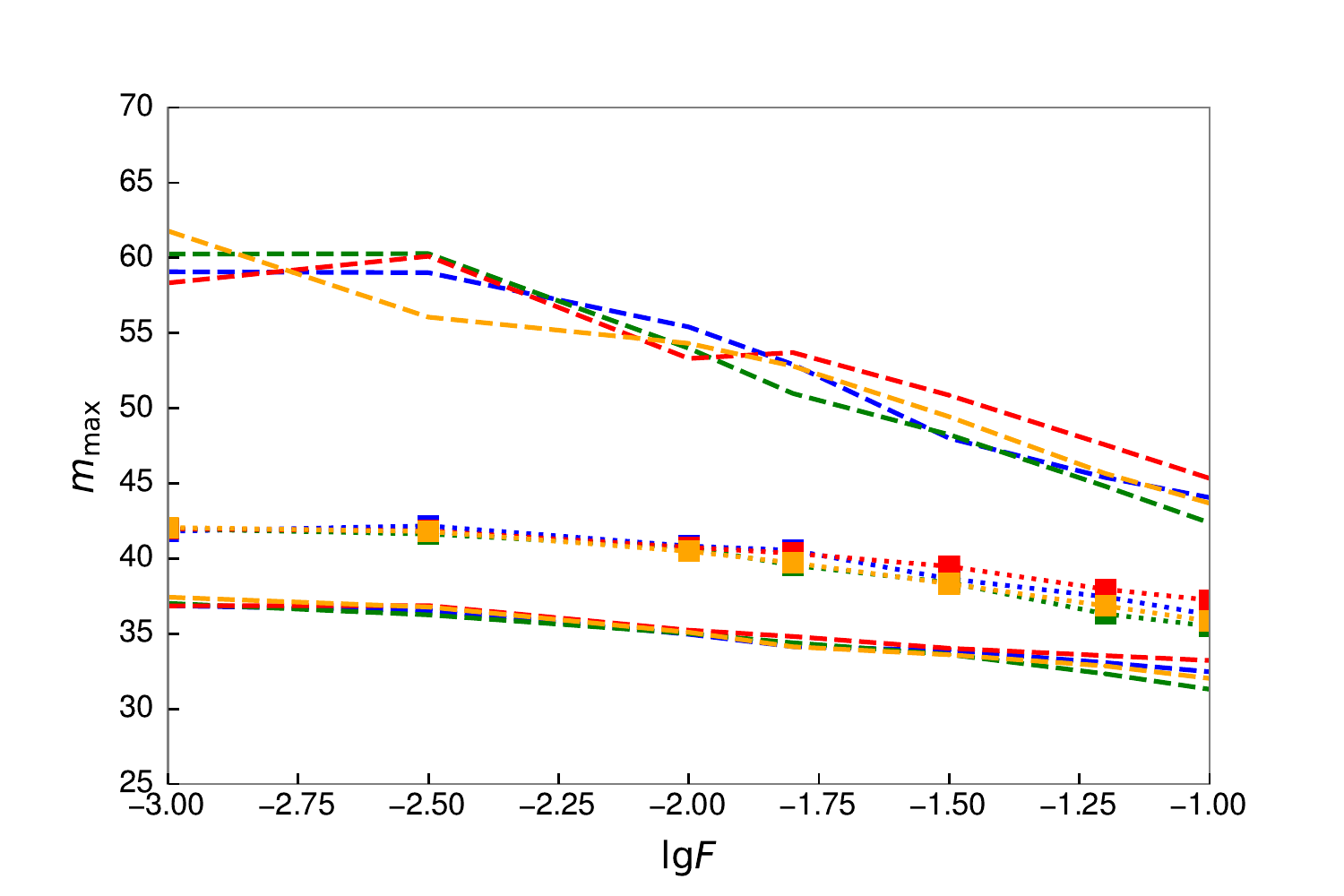}
	\includegraphics[angle=0,scale=0.5]{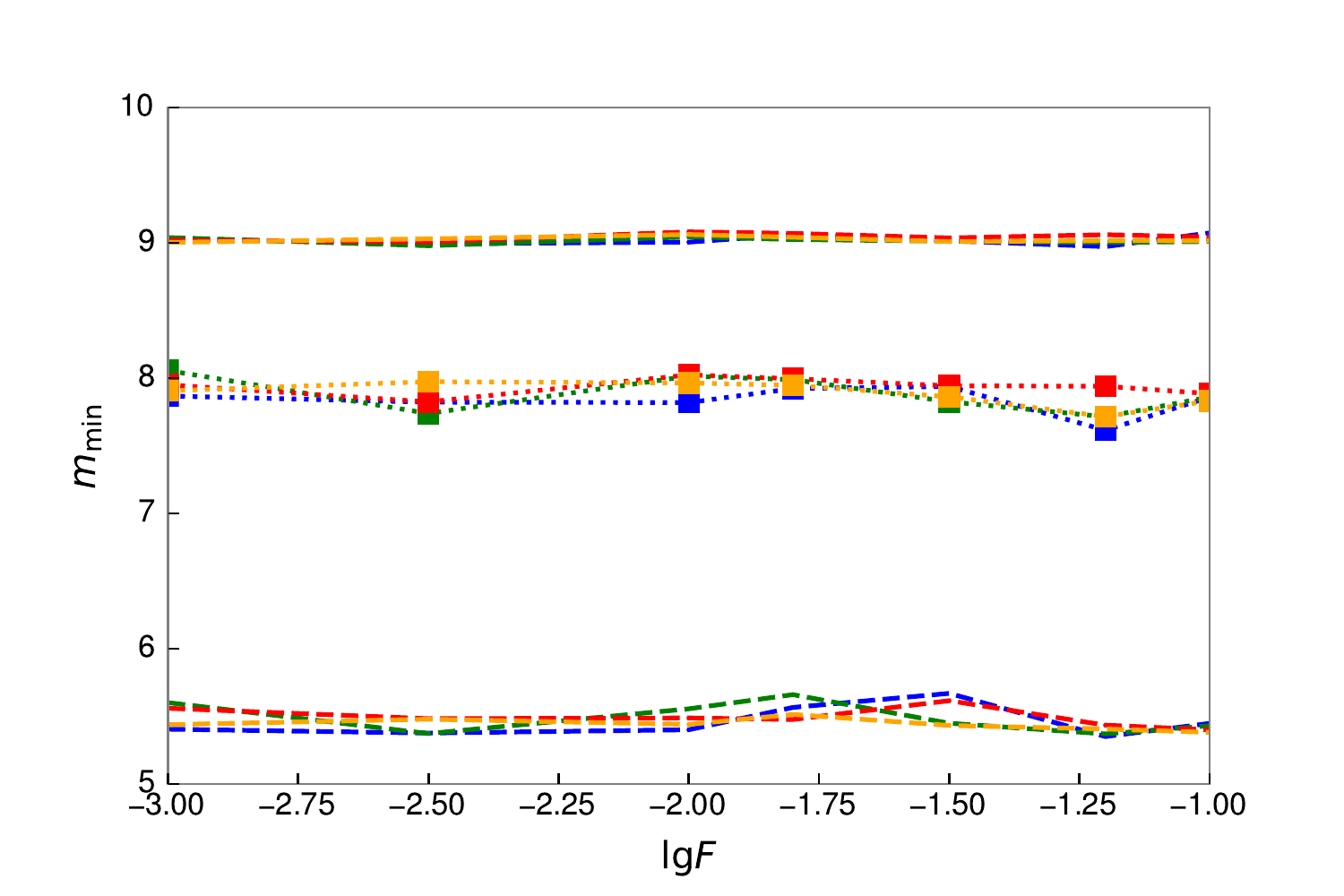}
	\includegraphics[angle=0,scale=0.5]{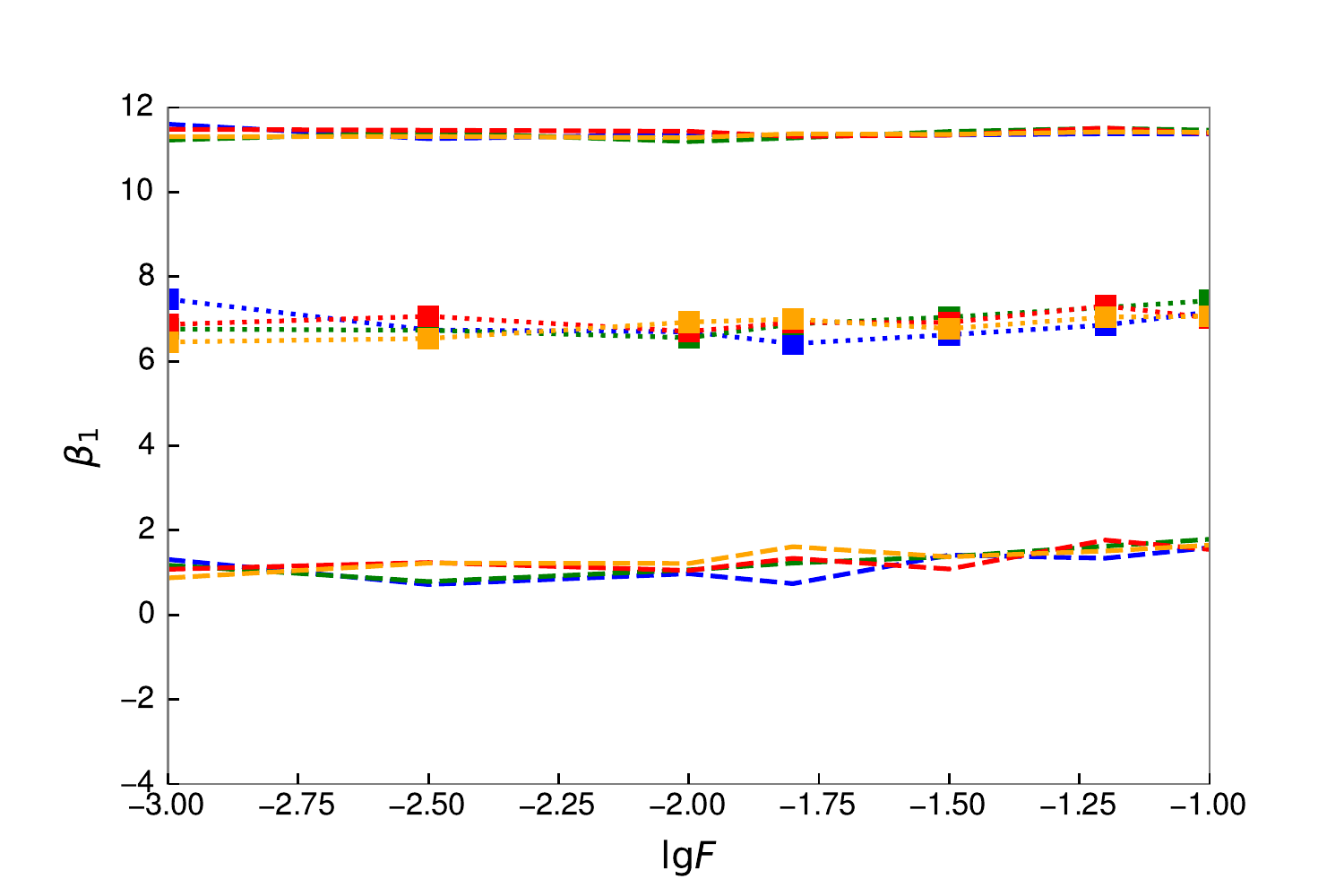}
	\caption{Dependency of the inferred parameters of Model 2PL on the choice of fixed parameters. The medians of the parameters are marked with filled squares, and the dashed lines show the $90\%$ credible intervals. Different choices of $\alpha_2$ and $m_{\rm edge}$ are represented by different colors. Blue: $\alpha_2=1,m_{\rm edge}=100$; green: $\alpha_2=2,m_{\rm edge}=100$; red:$\alpha_2=1,m_{\rm edge}=150$; orange: $\alpha_2=2,m_{\rm edge}=150$.}
	\hfill
\end{figure}

In this section, we show the results of hierarchical inference described above, and the credible intervals are reported with 90\% uncertainties unless otherwise stated. The constraints on the parameters of each models are reported in Tab.\ref{tb:results}. The result of Model B in our inference consist with that of \citet{2019ApJ...882L..24A}. The common parameters between Model B and Model Bcut are consistent with each other. The ``cut-off index" $k$ in Model Bcut has a large median value of 7.48, which indicates that the BHMF, if its main component has a power-law shape, does have a very hard cut-off after a certain limit of mass. In Fig.\ref{fig:depend}, we show the dependencies of the inferred $\alpha_1$, $m_{\rm max}$, $m_{\rm min}$ and $\beta_1$ on the choice of fixed values for Model 2PL. One can find from Fig.\ref{fig:depend} that the constraints on the parameters of the first segment are insensitive to the choice of $\alpha_2$ and $m_{\rm edge}$, while the constraints on $\alpha_1$ and $m_{\rm max}$ are sensitive to the choice of ${\rm lg}F$.

\begin{table*}[]
	\centering
	\caption{Summary of Constraints on the Parameters Considered in Tab.\ref{tb:priors}}
	\begin{ruledtabular}
		\label{tb:results}
		\begin{tabular}{c|ccc}
			Parameters/Models                     & Model B                     & Model Bcut                  & Model 2PL                     \\ \hline
			$\alpha$ or $\alpha_1$                & ${1.84}_{-1.61}^{+1.38}$    & ${1.57}_{-1.90}^{+1.56}$    & $(-1.06,3.22)$                \\
			$m_{\rm max}$ [$M_\odot$]                        & ${42.91}_{-5.46}^{+16.95}$  & N/A                         & $(31.30,61.78)$               \\
			$m_{\rm min}$ [$M_\odot$]                       & ${7.94}_{-2.56}^{+1.13}$    & ${7.81}_{-2.42}^{+1.21}$    & $(5.35,9.08)$                 \\
			$\beta_{\rm q}$ or $\beta_{\rm q,1}$  & ${6.91}_{-5.76}^{+4.64}$    & ${6.56}_{-5.48}^{+4.69}$    & $(0.72,11.60)$                \\
			$m_{\rm cut}$ [$M_\odot$]                        & N/A                         & ${41.80}_{-9.80}^{+22.83}$  & N/A                           \\
			$k$                                   & N/A                         & ${7.48}_{-4.72}^{+3.99}$    & N/A                           \\
			$\alpha_2$                            & N/A                         & N/A                         & fixed                         \\
			$\beta_{\rm q,2}$                     & N/A                         & N/A                         & 0                             \\
			$m_{\rm edge}$ [$M_\odot$]   & N/A                         & N/A                         & fixed                         \\
			${\rm lg}F$                           & N/A                         & N/A                         & fixed                         \\
		\end{tabular}
		\tablenotetext{}{The table shows the $90\%$ credible intervals of posterior distributions. The intervals for Model 2PL are obtained by combining the results of all cases.}
	\end{ruledtabular}
\end{table*}

\begin{figure}[]
	\figurenum{3}\label{fig:bf}
	\centering
	\includegraphics[angle=0,scale=0.8]{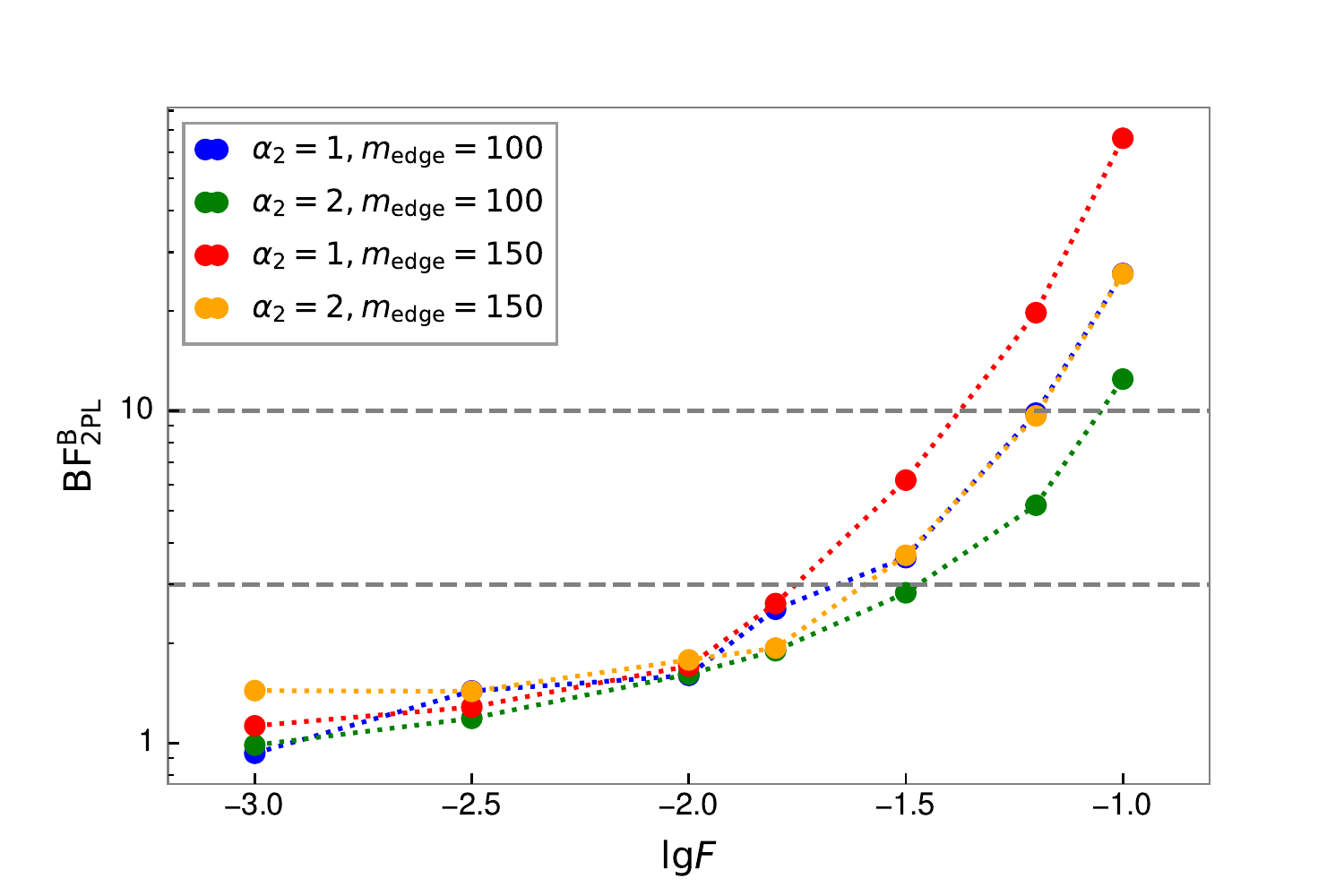}
	\caption{The Bayes factor between Model B and Model 2PL with different fix parameters of the second segment. The two gray dashed lines mark the ${\rm BF}^{\rm B}_{\rm 2PL}$ of 10 and 3, above which the evidence of ``Model B is more preferred by the data" is strong and moderate respectively.}
	\hfill
\end{figure}

Since Model 2PL can be regarded as Model B followed by an additional weak component, we investigate the possibility of the existence of such a weak component, by computing the Bayes factors ${\rm BF}^{\rm B}_{\rm 2PL}$ between the two models. We interpret Bayes factors of $< 3$ as anecdotal, $3-10$ as moderate, $10-30$ as strong, $30-100$ as very strong, and $> 100$ as decisive evidence for the first model is more favorable by the data compared to the second model. The results are displayed in Fig.\ref{fig:bf}, and it shows that ${\rm BF}^{\rm B}_{\rm 2PL}$ is most sensitive to the choice of ${\rm lg}F$. The ${\rm BF}$s in all groups are still lower than 10 even the ${\rm lg}F$ has reached $-1.5$ (corresponding to $R=0.032$). The ${\rm BF}$s in some groups are very close to or larger than 10 after the discrete values of ${\rm lg}F$ reaching $-1.2$ (correspinding to $R=0.063$). Accordingly, if we assume Model B is already capable of describing the main feature of current data, \textit{the evidence of excluding an extra segment of BHMF after $m_{\rm max}$ that extends to much higher mass is not strong enough, and this component is unlikely to occupy $>6\%$ of all primary BHs, but the fraction can still be as high as $3\%$.}

The resulting ${\rm BF}$ is affected by the choice of prior. In the analysis above, we have used relatively broad and un-informative priors for the parameters (see Tab.\ref{tb:priors} for details), and we expect that the information used in our model comparison mainly comes from the data. To study the influence from the choice of prior, we narrow down the volumes of the common priors for the four models (also shown in Table.\ref{tb:priors}) to recalculate the ${\rm BF}$s above, and find that the influence is relatively small: for example, the ${\rm BF}^{\rm B}_{\rm 2PL}$ with $(\alpha_2,m_{\rm edge}) = (1,100)$ is 2.17 for default prior volume, and 2.06 for the narrower prior volume; the ${\rm BF}^{\rm B}_{\rm 2PL}$ with $(\alpha_2,m_{\rm edge},{\rm lg}F) = (1,100,-1.5)$ is 4.88 for the default and 3.61 for the narrower prior volume (note that the ${\rm BF}$ derived from the nest sampling also has its uncertainty, which is $\sim 0.08$ and $\sim 0.05$ for the default and narrower prior volume respectively). As a consequence, our conclusion will still hold for un-informative priors with reasonably broader/narrower volumes.

\section{The chance of hearing the birth of IMBH in O3}\label{sec:imp}
In the above analysis, we showed that the fraction of high mass ($\gtrsim 40 M_\odot$) BHs can be as high as a few percents. Since the ground-based GW detectors is most sensitive to the BBHs with total mass $\sim 100-300\,M_\odot$, the chance of detecting IMBHs born from mergers can be promising. Previously, we have already studied this topic in \citet{2017arXiv170501881L}. However, since there were only a few BBH events being reported at that time, the BHMF took there needs to be modified now. Based on the O1-O2 data, we derive the expected detections and compare it with the current released information of O3. The Gravitational-Wave Candidate Event Database (GraceDB\footnote{\url{https://gracedb.ligo.org/superevents/public}}) reports a total number of 56 detection candidates, among which 38 events have the probability $\gtrsim 90\%$ to be BBHs (and 4 additional events have probability $\ge 95\%$ lie within the lower mass gap). GW190521 is confirmed to be a merger yielding IMBH, while the identification and properties of other BBH candidates are yet to be published. In the following we assume these 38 candidates are all BBHs.

\begin{figure}[]
	\figurenum{4}\label{fig:o3mass}
	\centering
	\includegraphics[angle=0,scale=0.5]{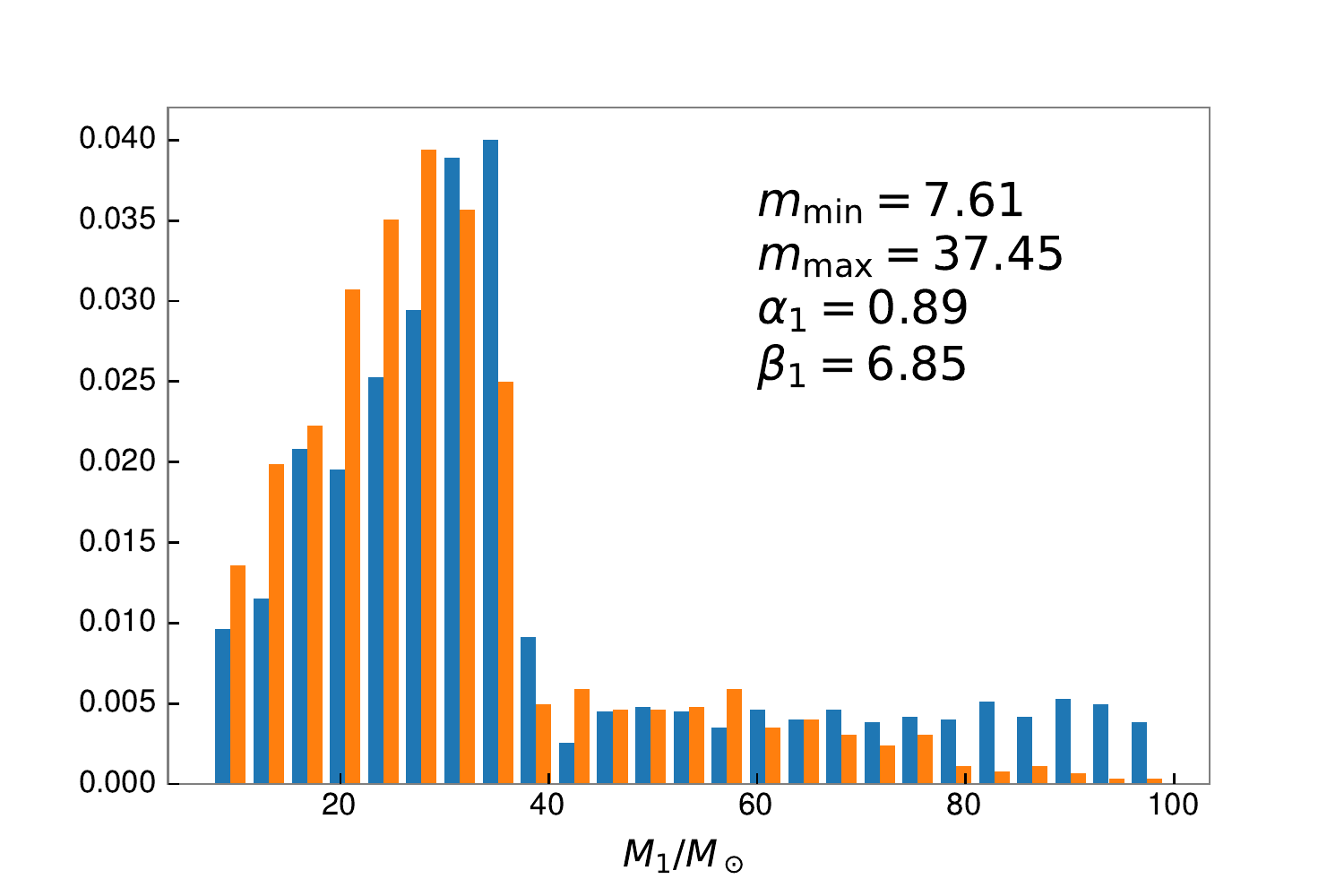}
	\includegraphics[angle=0,scale=0.5]{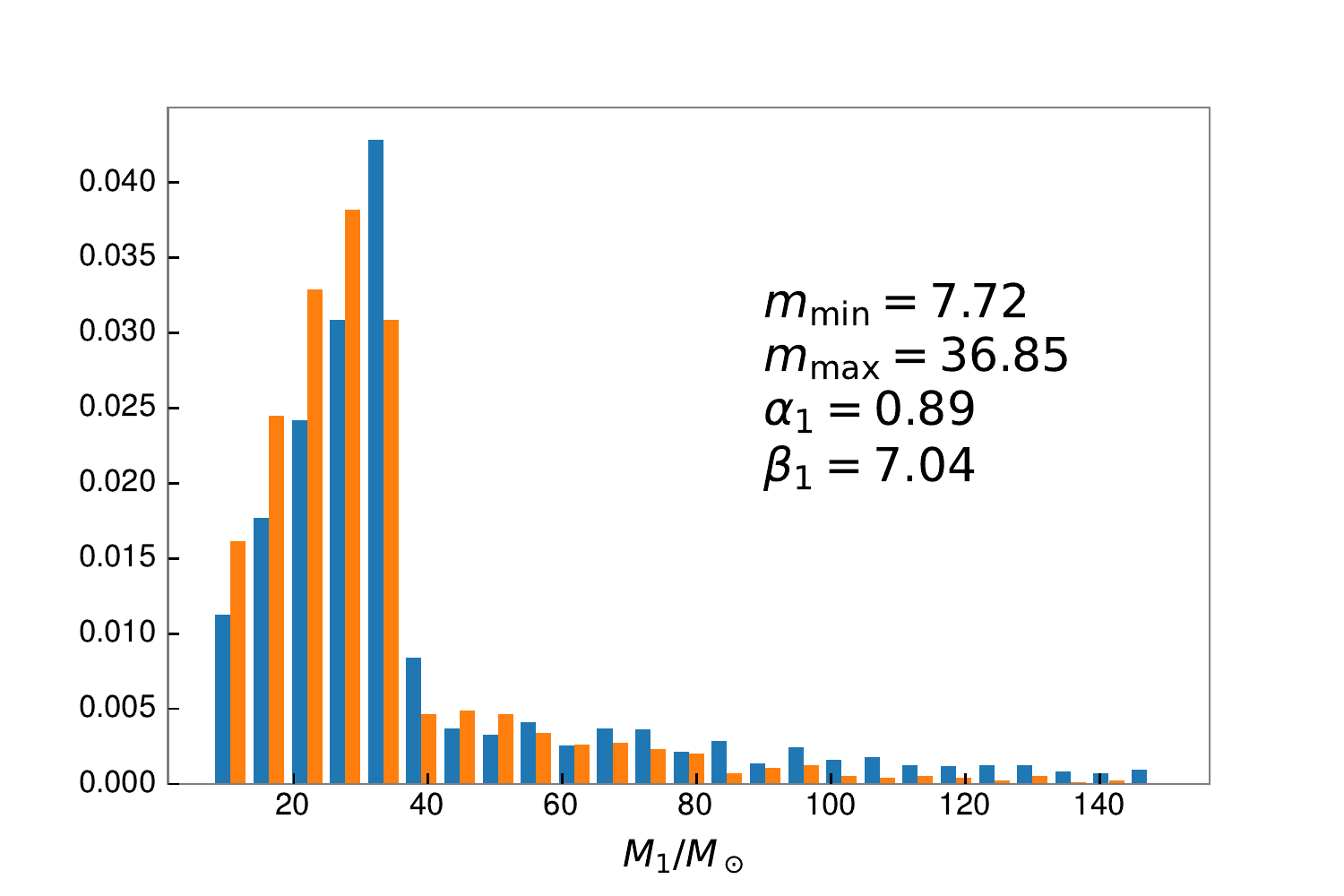}
	\includegraphics[angle=0,scale=0.5]{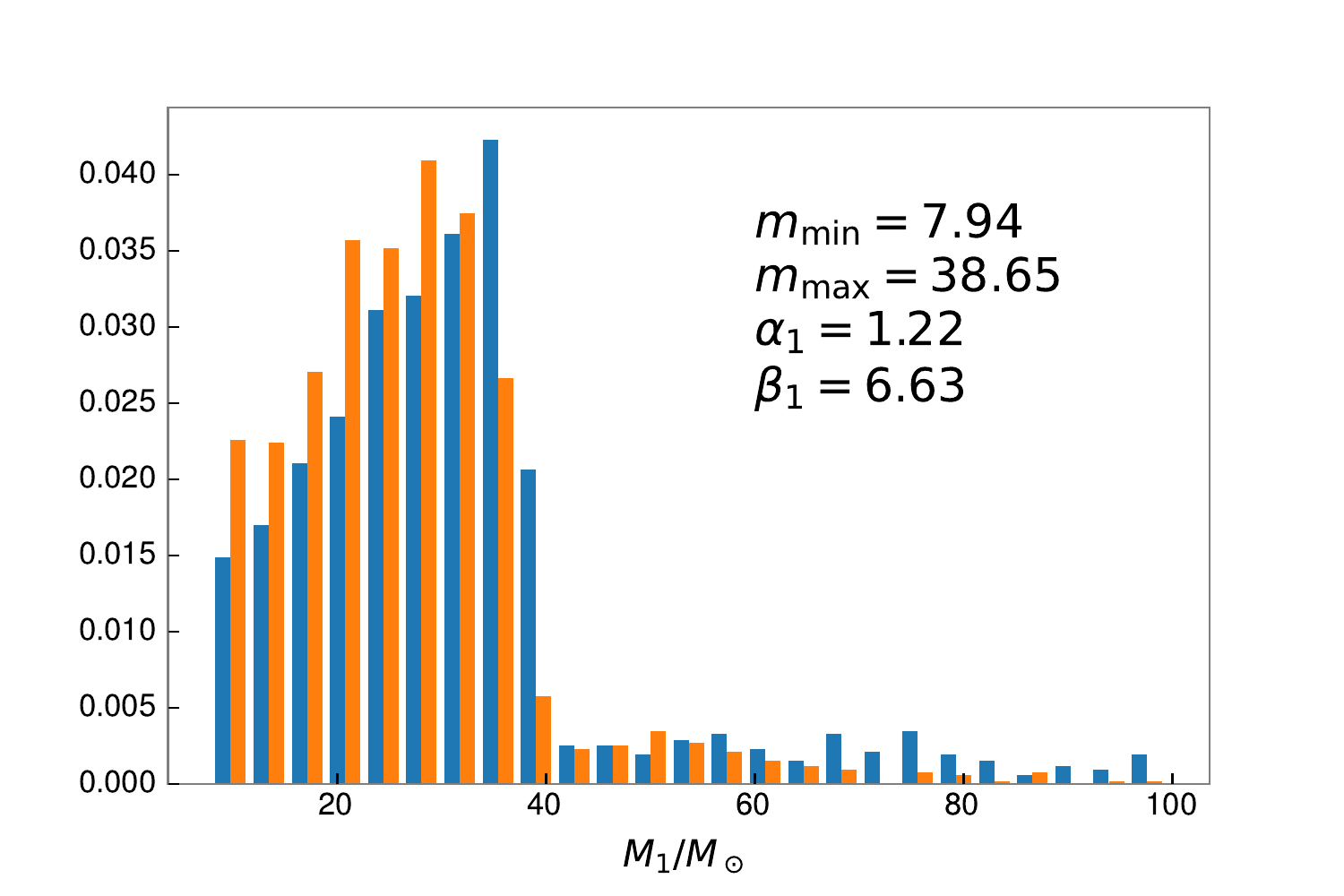}
	\includegraphics[angle=0,scale=0.5]{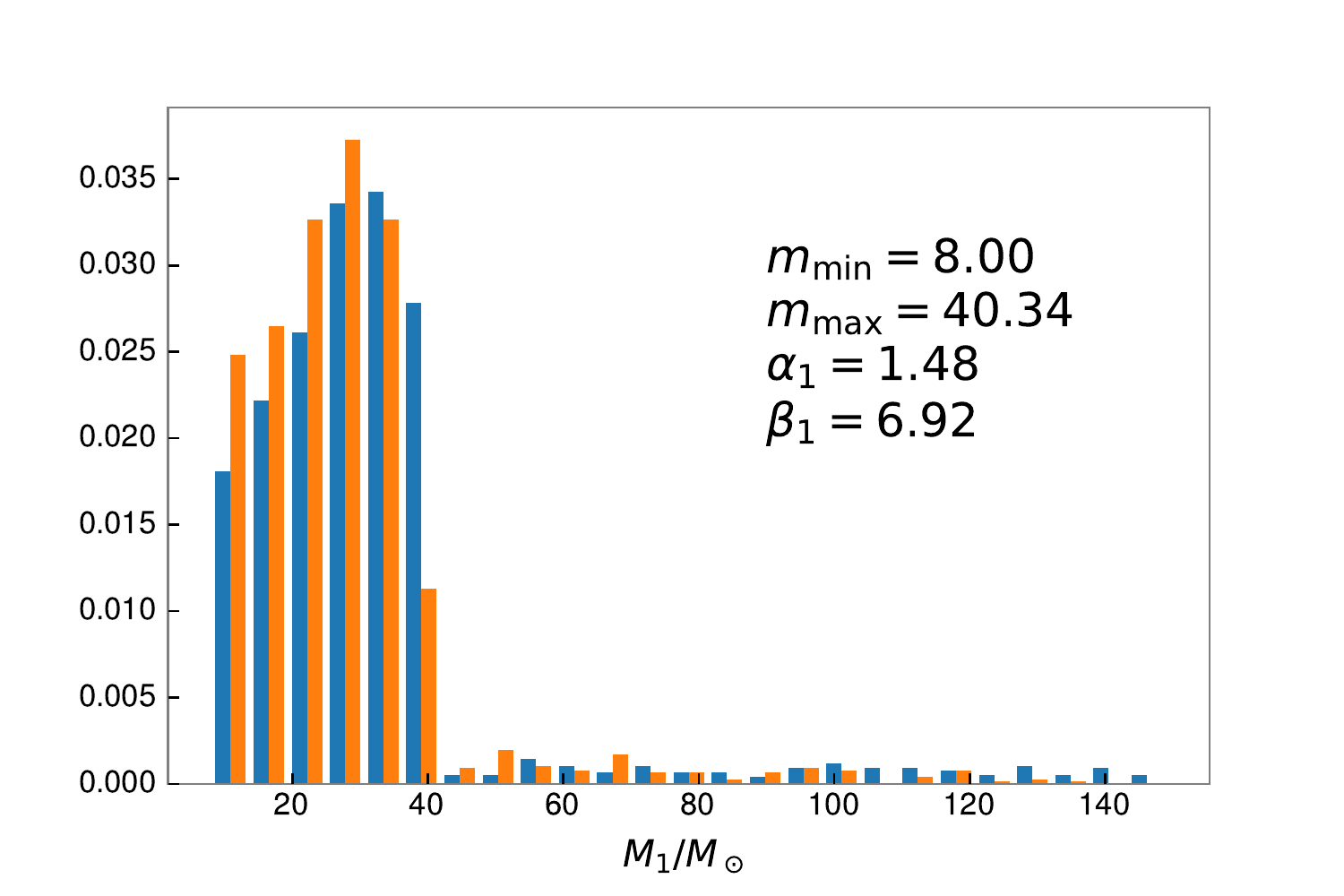}
	\caption{The observed primary (blue bars) and secondary (orange bars) mass distribution in LIGO/Virgo's O3 run predicted by Model 2PL with fix parameters. The parameters of the 2nd segment are selected by the Bayes factor criterion discussed in Fig.\ref{fig:bf}, and represent the most optimistic cases in our analysis for observing IMBH. The cases in the upper row are selected by ${\rm BF}^{\rm B}_{\rm 2PL} < 10$, and $\sim 16\%$ of their remnants are IMBHs; the cases in the lower row are selected by ${\rm BF}^{\rm B}_{\rm 2PL} < 3$, and $\sim 7\%$ of their remnants are IMBHs. Top left: the case with $m_{\rm edge} =100,\alpha_2=1$ and ${\rm lg}F=-1.2$; top right: the case with $m_{\rm edge} =150,\alpha_2=2$ and ${\rm lg}F=-1.2$; bottom left: the case with $m_{\rm edge} =100,\alpha_2=2$ and ${\rm lg}F=-1.5$; bottom right: the case with $m_{\rm edge} =150,\alpha_2=1$ and ${\rm lg}F=-1.8$. The parameters of the first segment is the one with the highest likelihood in the Bayes inference.}
	\hfill
\end{figure}

Among all the parameter sets presented in Fig.\ref{fig:bf}, ${\rm lg}F$ can be as large as $-1.2$ before the ${\rm BF}$ reaching 10 (the ``strong evidence of exclusion") for three groups, $(\alpha_2,m_{\rm edge}) = (2,100)$, $(\alpha_2,m_{\rm edge}) = (1,100)$, and $(\alpha_2,m_{\rm edge}) = (2,150)$. We consider the BHMF with parameters in the later two groups as the most optimal ones for observing the formation of IMBH. For each of the two groups, we simulate 100,000 events in which $(m_1,m_2)$ are drawn from population described by the parameters with the highest likelihood in the inference. Then we count the fraction of mergers with SNR$\ge 8$ (using the \textit{aLIGOLateLowSensitivity} PSD in {\sc PyCBC}) and $M_{\rm tot} > 100/0.95\,M_{\odot}$ (considering approximately $5\%$ of the total gravitational mass is radiated by GW), and find that the fraction of mergers forming IMBH can be as high as $\sim 16\%$ for both groups, which means that 6 out of the ``actual'' 38 detections have IMBH remnants. A more conservative case may be accessed by choosing the parameter sets with ${\rm BF}<3$. By applying this criteria, the most optimistic expectation on the fraction of mergers that form IMBH is $\sim 7\%$ (for the BH population with $(\alpha_2,m_{\rm edge},{\rm lg}F) = (2,100,-1.5)$ and $(\alpha_2,m_{\rm edge},{\rm lg}F) = (1,150,-1.8)$), corresponding to about 3 out of the 38 detections. We show the observed distribution of BH masses predicted by the BHMF of these most optimistic cases in Fig.\ref{fig:o3mass}. Note that these are the expected mean values of IMBH detections, and due to poisson fluctuation the observed numbers have variances equal to the mean values. It is worthy of pointing out that although the above expectations for IMBH detections are similar for $m_{\rm edge}=100$ and $m_{\rm edge}=150$, the population with $\alpha_2=1$, $m_{\rm edge}=150$ and ${\rm lg}F=-1.8$ (and ${\rm BF}<3$) predicts $3\%$ (1 out of 38) of its mergers having $m_1>100\,M_\odot$. This implies that we have the chance to clarify whether a single pre-merger star can be the lightest IMBH (with a mass $\geq 100\,M_\odot$) in O3.

\section{Conclusion and Discussion}\label{sec:dis}
In this work, we have studied the BHMFs based on the 10 BBHs observed in LIGO's O1-O2 runs, and considered three models presented in Tab.\ref{tb:priors}. By performing Bayesian hierarchical inference, we find that the data do prefer a very sharp cut-off at the mass $\sim 40 M_{\rm \odot}$ in the BHMF. However, there is still a good fraction of parameter space in which the model with an extra power-law segment can not be ruled our with strong evidence by the single power-law model. The fraction of the extra segment to the overall distribution 
can be high up to $3\%$. Such a new segment can give rise to rather exciting results in LIGO/Virgo's O3 and future observation runs. The fraction of mergers that form IMBHs can be as large as $\sim 16\%$ in the most optimistic case and can be $\sim 7\%$ in a more conservative case. Thus, the discovery of GW190521 is within our expectation.

There are space of improvement in the future works. In our analysis, we only include the information from the mass distributions inferred from each event. The inclusion of spin data (although additional thoughts are needed to construct the spin model) would enhance the ability for model comparison in the inference. For example, the expected spins in the second segment of Model 2PL could be larger and more isotropic if we assume the BBHs in the first segment is formed by field binary stars and the second segment is formed by dynamical capture. It is worthy to note that the most heavy BBH GW170729 (with source frame primary mass of $\sim 50 M_{\odot}$) has $\chi_{\rm eff} \sim 0.37$ \citep{2019PhRvX...9c1040A}, while the GW190521 has $\chi_{\rm p} \sim 0.6$ \citep{2020arXiv200901190T}, which are larger than the spins of other BBH events with lower masses. Limited by the number of data, the investigation on the parameter space of Model 2PL is restricted in this work. With several tens of new events in O3, especially the high mass event like GW190521, the properties of the second/high-mass segment can be better constrained.

Finally, we would like to remind that the 2PL BHMF model is still very likely to be a too simplified approximation of the real scenario. In particular, at the masses $\geq 133M_\odot$, the BHMF may show another interesting structure. On the one hand, the supernova explosions can directly produce such massive black holes \citep{2002RvMP...74.1015W}. On the other hand, the mergers of some black hole binaries can also produce IMBHs with the masses $\geq 133M_\odot$, as already observed in GW190521 \citep{2020arXiv200901075T}. All these objects could be involved in the BBH mergers and might be detected by the advanced LIGO/Virgo in the next decade.

\section{Acknowledgment}
We thank Dr. Y. M. Hu for kind help. This work was supported in part by NSFC under grants of No. 11525313 (i.e., Funds for Distinguished Young Scholars), No. 11921003 and No. 11933010, the Funds for Distinguished Young
Scholars of Jiangsu Province (No. BK20180050), the Chinese Academy of Sciences via the Strategic Priority Research Program (Grant No. XDB23040000), Key Research Program of Frontier Sciences (No. QYZDJ-SSW-SYS024). This research has made use of data and software obtained from the Gravitational Wave Open Science Center (https://www.gwopenscience.org), a service of LIGO Laboratory, the LIGO Scientific Collaboration and the Virgo Collaboration. LIGO is funded by the U.S. National Science Foundation. Virgo is funded by the French Centre National de Recherche Scientifique (CNRS), the Italian Istituto Nazionale della Fisica Nucleare (INFN) and the Dutch Nikhef, with contributions by Polish and Hungarian institutes.

\software{Bilby \citep[version 0.6.9, ascl:1901.011, \url{https://git.ligo.org/lscsoft/bilby/}]{2019ascl.soft01011A}, PyCBC \citep[version 1.13.6, ascl:1805.030, \url{https://github.com/gwastro/pycbc}]{2018ascl.soft05030T}, PyMultiNest \citep[version 2.6, ascl:1606.005, \url{https://github.com/JohannesBuchner/PyMultiNest}]{2016ascl.soft06005B}}

\end{document}